\documentclass[12pt]{article}
\usepackage{epsfig}
\newcommand{\as}{\alpha_s}
\newcommand{\ord}{{\cal O}}

\newcommand{\lamps}{\lambda_p^{(s)}}

\begin{document}


\thispagestyle{empty}
\begin{flushright}
CERN-TH/2001-151\\
MPI-PHT-2001-16\\
hep-ph/0106081\\
May 2001
\end{flushright}

\vspace*{1.5cm}
\centerline{\Large\bf The Radiative Decays $B\to V\gamma$}
\vspace*{0.3cm}
\centerline{\Large\bf at Next-to-Leading Order in QCD}
\vspace*{2cm}
\centerline{{\sc Stefan W. Bosch}\ ${}^{a,b}$ and 
{\sc Gerhard Buchalla}\ ${}^a$}
\bigskip
\bigskip
\centerline{\sl ${}^a$ Theory Division, CERN, CH-1211 Geneva 23,
                Switzerland}
\centerline{\sl ${}^b$ Max-Planck-Institut f\"ur Physik,
 Werner-Heisenberg-Institut,}
\smallskip
\centerline{\sl F\"ohringer Ring 6, D-80805 Munich, Germany}

\vspace*{2.5cm}
\centerline{\bf Abstract}
\vspace*{0.3cm}
We provide a model-independent framework for the analysis
of the radiative $B$-meson decays $B\to K^*\gamma$ and $B\to\rho\gamma$.
In particular, we give a systematic discussion of the various 
contributions to these exclusive processes based on the heavy-quark
limit of QCD. We propose a novel factorization formula for the consistent
treatment of $B\to V\gamma$ matrix elements involving charm (or up-quark)
loops, which contribute at leading power in $\Lambda_{QCD}/m_B$ to
the decay amplitude. Annihilation topologies are shown to be power suppressed.
In some cases they are nevertheless calculable. 
The approach is similar to the framework of QCD factorization
that has recently been formulated for two-body non-leptonic $B$ decays.
These results allow us, for the first time, to compute 
{\it exclusive\/} $b\to s(d) \gamma$ decays systematically beyond
the leading logarithmic approximation. We present 
results for these decays complete to next-to-leading order in QCD
and to leading order in the heavy-quark limit.
Phenomenological implications for various observables of interest
are discussed, including direct CP violation, and isospin and U-spin
breaking effects.

\noindent 
\vfill

\begin{flushleft}
CERN-TH/2001-151
\end{flushleft}
 
\newpage
\pagenumbering{arabic}

\section{Introduction}\label{sec:intro}

The radiative transitions $b\to s\gamma$, $b\to d\gamma$ are among
the most valuable probes of flavour physics.
Proceeding at rates of order $G^2_F\alpha$, they are
systematically enhanced over other loop-induced, non-radiative
rare decays, which are proportional to $G^2_F\alpha^2$.
In fact, the Cabibbo-favoured $b\to s\gamma$ modes belong to the
small number of rare decays that are experimentally accessible
already at present.
The inclusive branching fraction has been measured to be
\begin{equation}\label{bsgamex}
B(B\to X_s\gamma)=(2.96\pm 0.35)\cdot 10^{-4}
\end{equation}
combining the results of \cite{BLA,TAY,BAR}.
The branching ratios for the exclusive channels have been determined
by CLEO \cite{COA}, 
and more recently also by BABAR \cite{PUL} and BELLE \cite{TAY}:
\begin{equation}\label{b0kgamex}
B(B^0\to K^{*0}\gamma)=\left\{
\begin{array}{ll}
(4.55\pm 0.70\pm 0.34)\cdot 10^{-5} & \cite{COA} \\
(4.39\pm 0.41\pm 0.27)\cdot 10^{-5} & \cite{PUL} \\
(4.96\pm 0.67\pm 0.45)\cdot 10^{-5} & \cite{TAY} 
\end{array}\right.
\end{equation}
\begin{equation}\label{bpkgamex}
B(B^+\to K^{*+}\gamma)=\left\{
\begin{array}{ll}
(3.76\pm 0.86\pm 0.28)\cdot 10^{-5} & \cite{COA} \\
(3.89\pm 0.93\pm 0.41)\cdot 10^{-5} & \cite{TAY} 
\end{array}\right.
\end{equation}
On the theoretical side, the flavour-changing neutral current (FCNC)
reactions $b\to s(d)\gamma$ are characterized by their high sensitivity
to New Physics 
and by the particularly large impact of short-distance
QCD corrections \cite{BBM,AGM,BMMP,CFMRS}.
Considerable efforts have therefore been devoted to achieve a full
calculation of the inclusive decay $b\to s\gamma$ at next-to-leading
order (NLO) in renormalization group (RG) improved perturbation
theory \cite{AY,GHW,CMM} (see \cite{MG} for recent reviews).

Whereas the inclusive mode can be computed perturbatively, using
the fact that the $b$-quark mass is large and employing the heavy-quark
expansion, the treatment of the exclusive channel $B\to K^*\gamma$
is in general more complicated. In this case bound state effects
are essential and need to be described by nonperturbative 
hadronic quantities (form factors).
The basic mechanisms at next-to-leading order were already
discussed previously for the $B\to V\gamma$ amplitudes \cite{AAWGSW}.
However, hadronic models were used to evaluate the various
contributions, which did not allow a clear separation
of short- and long-distance dynamics and a clean distinction
of model-dependent and model-independent features.

In this paper we present a systematic analysis of the exclusive
radiative decays $B\to V\gamma$ ($V=K^*$, $\rho$) in QCD, based
on the heavy quark limit $m_b\gg\Lambda_{QCD}$. We shall establish
factorization formulas for the evaluation of the relevant hadronic
matrix elements of local operators in the weak Hamiltonian.
Factorization holds in QCD to leading power in the heavy quark limit.
This result relies on arguments similar to those used previously
to demonstrate QCD factorization for hadronic two-body modes
of the type $B\to\pi\pi$ \cite{BBNS1,BBNS2}.

This framework will allow us to separate perturbatively calculable
contributions from the nonperturbative form factors and universal meson
light-cone distribution amplitudes (LCDA) in a systematic way. 
This includes the treatment of loop effects from light quarks,
in particular up and charm.
Such loop effects are straightforwardly included for the inclusive
decays $b\to s(d)\gamma$. For the exclusive modes, however,
the effects from virtual charm and up quarks have so far been
considered to be uncalculable ``long-distance'' contributions
and have never been treated in a model independent fashion.

Finally, power counting in $\Lambda_{QCD}/m_b$  implies a hierarchy
among the possible mechanisms for $B\to V\gamma$ transitions. This
allows us to identify leading and subleading contributions.
For example, weak annihilation contributes only at subleading
power in the heavy quark limit.

Within this approach, higher order QCD corrections can be consistently
taken into account. We give the $B\to V\gamma$ decay amplitudes
at next-to-leading order (NLO).
After including NLO corrections the largest uncertainties still come
from the $B\to V$ form factors, which are at present known only with 
limited precision ($\sim \pm 15\%$), mostly from QCD
sum rule calculations \cite{BB98}.
The situation should improve in the future with the help of
both lattice QCD \cite{FS} and analytical methods based on the
heavy-quark and large-energy limits \cite{CLOPR,BF,BH}.

Despite the more complicated theoretical situation of the
exclusive channels in comparison to the inclusive decays, the
goal of obtaining a better understanding of the exclusive modes
is very well motivated. This is because the exclusive decays are
easier to investigate experimentally, especially in the difficult
environment of hadron machines as the Fermilab Tevatron or the
LHC at CERN. In any case the systematic uncertainties, both
experimental and theoretical, are very different for $B\to V\gamma$
and $b\to s(d)\gamma$. A careful study of the exclusive modes
can therefore yield valuable complementary information in testing
the Standard Model.

The remainder of this paper is organized as follows.
In section \ref{sec:basics} we collect basic expressions and present the structure
of the QCD factorization formulas for $B\to V\gamma$ matrix elements
in general terms. Sections \ref{sec:BKgam} and \ref{sec:Brhogam} give a detailed discussion of
$B\to K^*\gamma$ and $B\to\rho\gamma$, respectively.
Phenomenological applications are studied in section \ref{sec:phen}. Finally,
section \ref{sec:concl} contains our conclusions.

\section{Basic Formulas}\label{sec:basics}

The effective Hamiltonian for $b\to s\gamma$ transitions reads
\begin{equation}\label{heff}
{\cal H}_{eff}=\frac{G_F}{\sqrt{2}}\sum_{p=u,c}\lamps
\left[ C_1 Q^p_1 + C_2 Q^p_2 +\sum_{i=3,\ldots ,8} C_i Q_i\right]
\end{equation}
where
\begin{equation}\label{lamps}
\lamps=V^*_{ps}V_{pb}
\end{equation}
The operators are given by
\begin{eqnarray}\label{q1def}
Q^p_1 &=& (\bar sp)_{V-A}(\bar pb)_{V-A} \\
Q^p_2 &=& (\bar s_i p_j)_{V-A}(\bar p_j b_i)_{V-A} \\
Q_3 &=& (\bar sb)_{V-A} \sum_q (\bar qq)_{V-A} \\
Q_4 &=& (\bar s_i b_j)_{V-A} \sum_q (\bar q_j q_i)_{V-A} \\
Q_5 &=& (\bar sb)_{V-A} \sum_q (\bar qq)_{V+A} \\
Q_6 &=& (\bar s_i b_j)_{V-A} \sum_q (\bar q_j q_i)_{V+A} \\
\label{q7def}
Q_7 &=& \frac{e}{8\pi^2}m_b\, 
        \bar s_i\sigma^{\mu\nu}(1+\gamma_5)b_i\, F_{\mu\nu}\\
\label{q8def}
Q_8 &=& \frac{g}{8\pi^2}m_b\, 
        \bar s_i\sigma^{\mu\nu}(1+\gamma_5)T^a_{ij} b_j\, G^a_{\mu\nu}
\end{eqnarray}
The most important operators are $Q^p_{1,2}$ and the magnetic penguin
operator $Q_7$. Note that in our notation the numbering of $Q^p_{1,2}$
is reversed with respect to the convention of \cite{BBL}, 
i.e. $C_1^{(0)}(M_W)=1$ and $C_2^{(0)}(M_W)=0$.
The sign conventions for the electromagnetic and 
strong couplings correspond to the covariant derivative
$D_\mu=\partial_\mu +ie Q_f A_\mu + i g T^a A^a_\mu$. With these
definitions the coefficients $C_{7,8}$ are negative in the
Standard Model, which is the choice generally adopted
in the literature. 
The effective Hamiltonian for $b\to d\gamma$ is obtained from
(\ref{heff}--\ref{q8def}) by the replacement $s\to d$. 
The Wilson coefficients $C_i$ in (\ref{heff}) are known at
next-to-leading order \cite{CMM}.

The most difficult step in computing the decay amplitudes is the
evaluation of the hadronic matrix elements of the operators in 
(\ref{heff}). A systematic treatment can be given in the heavy-quark
limit. We will argue that in this case the following 
factorization formula is valid 
\begin{equation}\label{fform}
\langle V\gamma(\epsilon)|Q_i|\bar B\rangle =
\left[ F^{B\to V}(0)\, T^I_{i} +
\int^1_0 d\xi\, dv\, T^{II}_i(\xi,v)\, \Phi_B(\xi)\, \Phi_V(v)\right]
\cdot\epsilon
\end{equation}
where $\epsilon$ is the photon polarization 4-vector.
Here $F^{B\to V}$ is a $B\to V$ transition form factor,
and $\Phi_B$, $\Phi_V$ are leading twist light-cone distribution amplitudes
of the $B$ meson and the vector meson $V$, respectively.
These quantities are universal, nonperturbative objects. They
describe the long-distance dynamics of the matrix elements, which
is factorized from the perturbative, short-distance interactions
expressed in the hard-scattering kernels $T^I_{i}$ and $T^{II}_i$.
The QCD factorization formula (\ref{fform}) holds up to
corrections of relative order $\Lambda_{QCD}/m_b$.

\begin{figure}[t]
   \epsfysize=4cm
   \epsfxsize=6cm
   \centerline{\epsffile{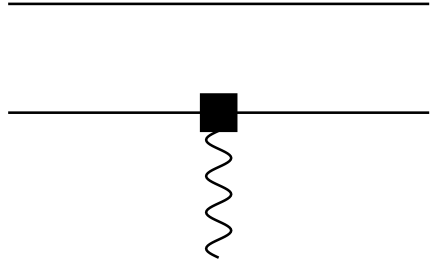}}
\caption{Contribution of the magnetic penguin operator $Q_7$
described by $B\to V$ form factors. All possible gluon exchanges 
between the quark lines are included in the form factors and have not
been drawn explicitly. \label{fig:q7}}
\end{figure}

For $Q_7$ the factorization formula (\ref{fform}) is trivial.
The matrix element is simply expressed in terms of the standard
form factor, $T^I_{7}$ is a purely kinematical function and
the spectator term $T^{II}_7$ is absent. An illustration is given
in Fig. \ref{fig:q7}.
The matrix element reads
\begin{equation}\label{q7f1f2}
\langle V(k,\eta) \gamma(q,\epsilon)|Q_7|\bar B\rangle =
-\frac{e}{2\pi^2}m_b\, c_V F_V \left[ 
\varepsilon^{\mu\nu\lambda\rho}\epsilon_\mu\eta_\nu k_\lambda q_\rho +
i (\epsilon\cdot\eta\, k\cdot q-\epsilon\cdot k\, \eta\cdot q)\right]
\end{equation}
where $c_V=1$ for $V=K^*$, $\rho^-$ and $c_V=1/\sqrt{2}$ for
$V=\rho^0$.
The $\bar B\to V$ form factor $F_V$ is evaluated at momentum transfer
$q^2=0$. 
Our phase conventions coincide with those of \cite{BB98,ABS}.
In particular we have $F_V>0$, $\varepsilon^{0123}=-1$, and the
phases of $V$ (with flavour content $\bar qq'$) and $\bar B$ are such that
\begin{eqnarray}\label{fbfv}
\langle V(k,\eta)|\bar q\sigma_{\mu\nu}q'|0\rangle &=&
-i(\eta_\mu k_\nu-\eta_\nu k_\mu)f^\perp_V \\
\langle 0|\bar u\gamma_\mu\gamma_5 b|\bar B(p)\rangle &=& +i f_B p_\mu
\end{eqnarray}
with positive $f_B$, $f^\perp_V$. In the leading logarithmic approximation
(LO) and to leading power in the heavy-quark limit, $Q_7$ gives the
only contribution to the amplitude of $\bar B\to V\gamma$.

\begin{figure}[t]
   \epsfysize=4cm
   \epsfxsize=12cm
   \centerline{\epsffile{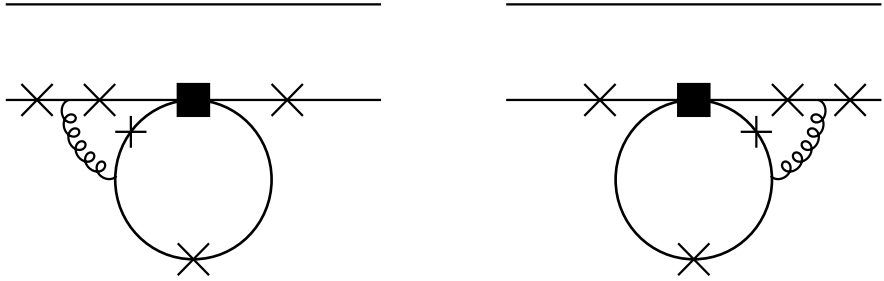}}
\caption{${\cal O}(\alpha_s)$ contribution to the hard-scattering 
kernels $T^I_{i}$ from four-quark operators $Q_i$.
The crosses indicate the places where the emitted photon can
be attached. \label{fig:qit1}}
\end{figure}

\begin{figure}[t]
   \epsfysize=2.5cm
   \epsfxsize=12cm
   \centerline{\epsffile{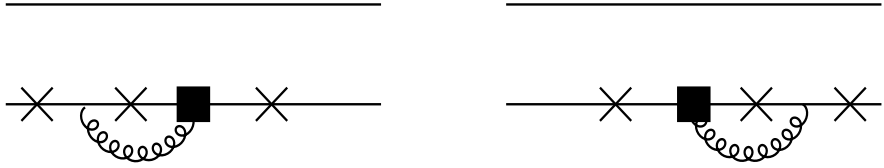}}
\caption{${\cal O}(\alpha_s)$ contribution to the hard-scattering 
kernels $T^I_{8}$ from chromomagnetic penguin operator $Q_8$.
\label{fig:q8t1}}
\end{figure}

\begin{figure}[t]
   \epsfysize=4cm
   \epsfxsize=15cm
   \centerline{\epsffile{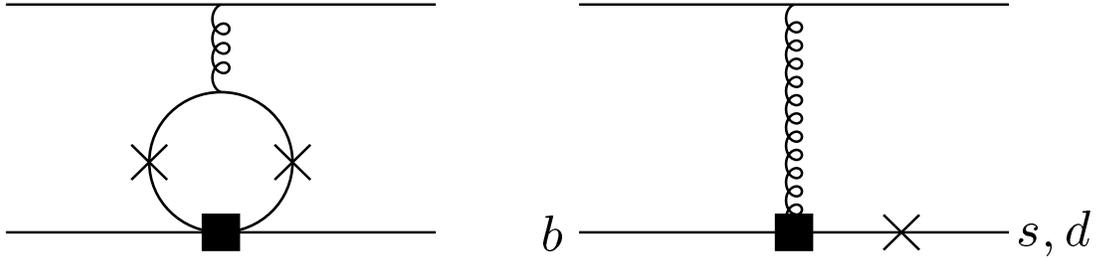}}
\caption{${\cal O}(\alpha_s)$ contribution to the hard-scattering 
kernels $T^{II}_i$ from four-quark operators $Q_i$ (left) and
from $Q_8$. \label{fig:qit2}}
\end{figure}

The matrix elements of the four-quark operators $Q_i$ (and of $Q_8$)
start contributing at ${\cal O}(\alpha_s)$. In this case the factorization
formula becomes nontrivial. The diagrams for the hard-scattering kernels
$T^I_{i}$ are shown in Fig. \ref{fig:qit1} 
for $Q_1,\ldots, Q_6$ and in Fig. \ref{fig:q8t1}
for $Q_8$. The non-vanishing contributions to
$T^{II}_i$ are shown in Fig.~\ref{fig:qit2}. 

The diagrams in Figs. \ref{fig:qit1} -- \ref{fig:qit2} represent
the complete set of contributions from $Q_{1,2}$ and $Q_8$
to leading order in the heavy-quark limit and at ${\cal O}(\alpha_s)$.
The first term in the factorization formula (\ref{fform}) is
given by the form factor from (\ref{q7f1f2}) and the hard-scattering
kernels $T^I_{i}$. The latter are calculable functions of
$m_q/m_b$, where $m_q$ is the mass of the internal quarks in the loop
diagram. For the second term in (\ref{fform}), where the spectator
quark is involved, we need light-cone distribution amplitudes
(wave functions) for $B$ mesons and light vector mesons.
In the case of the $B$ meson we have to leading power \cite{BBNS2,BF}
\begin{equation}\label{phi12def}
\langle 0|b(0) \bar u(z)|\bar B(p)\rangle =
\frac{i f_B}{4}(\not\! p + m_b)\gamma_5\,
\int^1_0 d\xi\, e^{-i\xi p_+ z_-}[\Phi_{B1}(\xi)+\not\! n \Phi_{B2}(\xi)]
\end{equation}
with $n=(1,0,0,-1)$, chosen to be parallel to the 4-momentum of
the vector meson. The functions $\Phi_{1,2}(\xi)$ describe
the distribution of light-cone momentum fraction $\xi=l_+/p_+$
of the spectator quark with momentum $l$ inside the $B$ meson.
Here light-cone components of four-vectors $v$ are defined by
\begin{equation}\label{vlcc}
v_\pm =\frac{v^0\pm v^3}{\sqrt{2}}
\end{equation}
The wave functions are highly asymmetric with 
$\xi={\cal O}(\Lambda_{QCD}/m_b)$. They are normalized as
\begin{equation}\label{phi12norm}
\int^1_0 d\xi\, \Phi_{B1}(\xi)=1 \qquad
\int^1_0 d\xi\, \Phi_{B2}(\xi)=0
\end{equation}
The first negative moment of $\Phi_{B1}(\xi)$, which will be needed below,
can be pa\-ra\-me\-trized by a quan\-ti\-ty 
$\lambda_B={\cal O}(\Lambda_{QCD})$, i.e.
\begin{equation}\label{lambdef}
\int^1_0 d\xi \frac{\Phi_{B1}(\xi)}{\xi}=\frac{m_B}{\lambda_B}
\end{equation}

In $\bar B\to V\gamma$ decays the vector meson is transversely
polarized. The leading-twist and leading-power distribution amplitude 
for light vector mesons with transverse polarization, $\Phi_\perp$, 
is defined by
\begin{equation}\label{phiperpdef}
\langle V(k,\eta)|q'(z)\bar q(0)|0\rangle =\frac{i f^\perp_V}{4}
\sigma^{\mu\nu}\eta_\mu k_\nu \int^1_0 dv\, e^{i\bar v k\cdot z}
\Phi_\perp(v)
\end{equation}
Here and in the following we use the short-hand notation
\begin{equation}
\bar v \equiv 1-v
\end{equation}
for light-cone variables.

The light-cone wave function $\Phi_\perp$ has an expansion in terms
of Gegenbauer polynomials $C^{3/2}_n(2v-1)$
\begin{equation}\label{phiperpgbp}
\Phi_\perp(v)=6v(1-v)\left[ 1+\sum^\infty_{n=1}
\alpha^\perp_n(\mu) C^{3/2}_n(2v-1)\right]
\end{equation}
where $C^{3/2}_1(x)=3x$, $C^{3/2}_2(x)=\frac{3}{2}(5x^2-1)$, etc.
The Gegenbauer moments $\alpha^\perp_n(\mu)$ are multiplicatively
renormalized. They vanish logarithmically as the scale $\mu\to\infty$.
In this limit $\Phi_\perp$ reduces to its asymptotic form
$\Phi_\perp(v)=6v\bar v$, which often is a reasonable first approximation.
The remaining leading-twist light-cone wave functions for light
vector mesons, $\Phi_\parallel$, $g_\perp^{(v)}$ and
$g_\perp^{(a)}$ \cite{BB98,BB96}, do not contribute at leading power
if the mesons are transversely polarized.

The form factor $F_V$, $f_B/\lambda_B$, and the light-cone wave function
$f^\perp_V \Phi_\perp(v)$ are the nonperturbative
quantities required to describe $\bar B\to V\gamma$ at next-to-leading
order in QCD. ($\Phi_{B2}(\xi)$ does not contribute.)

\begin{figure}[t]
   \epsfysize=4cm
   \epsfxsize=6cm
   \centerline{\epsffile{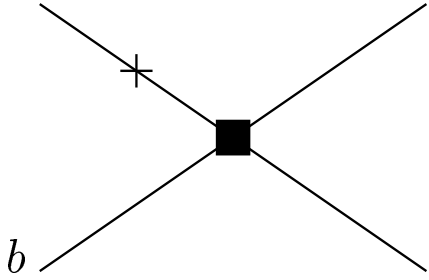}}
\caption{Annihilation contribution to $\bar B\to V\gamma$ decay.
The dominant mechanism is the radiation of the photon from
the light quark in the $B$ meson, as shown. This amplitude is
suppressed by one power of $\Lambda_{QCD}/m_b$, but it is still
calculable in QCD factorization. Radiation of the photon from the
remaining three quark lines is suppressed by $(\Lambda_{QCD}/m_b)^2$
for operators $Q_{1,2}$. \label{fig:ann}}
\end{figure}

There are further mechanisms that can in principle contribute
to $\bar B\to V\gamma$ decays.
One possibility is weak annihilation, depicted in Fig. \ref{fig:ann}.
In this case the leading-power projection onto the meson $V$
in (\ref{phiperpdef}) vanishes, because the trace over an odd
number of Dirac matrices is zero. A non-vanishing result arises
from the projector
\begin{equation}\label{fvmveta}
\langle V(k,\eta)| \bar q\gamma_\nu q' |0\rangle = f_V m_V \eta_\nu
\end{equation}
which, however, is suppressed by one power of $\Lambda_{QCD}/m_b$
compared to (\ref{fbfv}), (\ref{phiperpdef}) for transverse
polarization $\eta_\nu$ ($f_V$, $f^\perp_V$, $m_V\sim \Lambda_{QCD}$,
$k\sim m_b$).
This results in a corresponding power suppression of weak annihilation.
The dominant contribution comes from the diagram shown in
Fig. \ref{fig:ann}. Here the photon is emitted from the light-quark
constituent in the $B$ meson, which leads to a quark propagator
scaling as $1/\Lambda_{QCD}$. This is in contrast to the
remaining three possible diagrams with a quark propagator $\sim 1/m_b$.
The latter contributions are therefore even stronger suppressed,
$\sim (\Lambda_{QCD}/m_b)^2$ relative to the leading $\bar B\to V\gamma$
amplitude.

Despite its power suppression, the dominant annihilation amplitude
can be computed within QCD factorization. This is because the
colour-transparency argument applies to the emitted, highly energetic
vector meson in the heavy-quark limit \cite{BBNS2}.
A similar observation was already made in \cite{GP}.
On the other hand, we disagree with the claim made in this paper
that the ${\cal O}(\alpha_s)$ correction to Fig. \ref{fig:ann}
would vanish identically in the chiral limit and to leading-twist
order. This claim is based on the observation that the projector in
(\ref{phiperpdef}) gives zero when applied to a current with an odd 
number of Dirac matrices. However, by the same argument, the diagram
in Fig. \ref{fig:ann} would vanish even at leading order
in $\alpha_s$, which is not the case.
The proper treatment
of the ${\cal O}(\alpha_s)$ correction should employ the subleading-power 
projections related to the wave functions
$\Phi_\parallel$, $g_\perp^{(v)}$, $g_\perp^{(a)}$ \cite{BB98,BB96},
corresponding to the use of (\ref{fvmveta}) at ${\cal O}(\alpha^0_s)$.
This correction has not yet been computed, but we see no a-priori
reason to expect a vanishing result.

Since weak annihilation is a power correction, we will content 
ourselves with the lowest order result (${\cal O}(\alpha^0_s)$)
for our estimates below. In particular, we shall include the 
annihilation effects from operators $Q_{1,2}$ to estimate
isospin-breaking corrections in $B\to\rho\gamma$ decays.
The reason for including this class of power corrections is that they
come with a numerical enhancement from the large Wilson
coefficients $C_{1,2}$ ($C_1\approx 3|C_7|$) and are not CKM suppressed.
Instead, a CKM suppression of annihilation effects occurs for
$B\to K^*\gamma$ and these contributions are thus very small in this case.

\begin{figure}[t]
   \epsfysize=4cm
   \epsfxsize=15cm
   \centerline{\epsffile{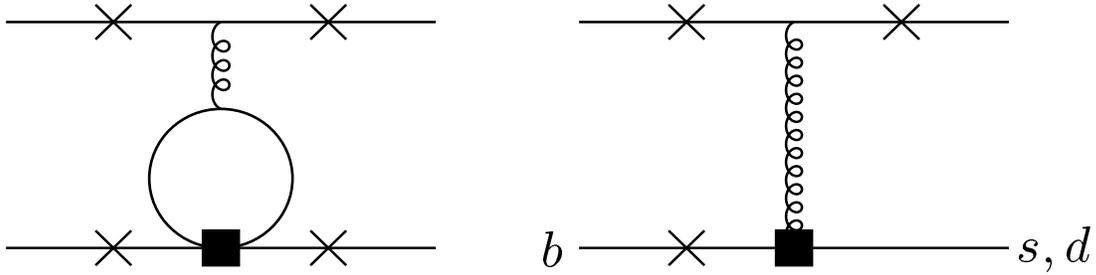}}
\caption{Other contributions that are power-suppressed in the
heavy-quark limit. \label{fig:subl}}
\end{figure}

Finally, Fig. \ref{fig:subl} displays the remaining diagrams
with a power suppression. 
The penguin-annihilation diagrams (left)
are power suppressed in a way similar to the ordinary annihilation
topologies. The spectator diagrams from $Q_8$ (right) lead to
amplitudes that are either manifestly power-suppressed
(photon emitted from the quark line in the upper right),
or that are superficially of leading power, but vanish when the
leading-order projections are performed (photon emitted from
either of the quarks forming the $B$ meson).
Recall, however, that photon emission from the light-quark line
from $Q_8$ is a leading-power effect (see Fig. \ref{fig:qit2}).

\section{$B\to K^*\gamma$}\label{sec:BKgam}

In the case of $B\to K^*\gamma$ the component of the
Hamiltonian (\ref{heff}) proportional to $\lambda_u$ is strongly
CKM suppressed ($|\lambda_u/\lambda_c|\approx 0.02$) and has
only a minor impact on the decay rate. It is essentially
negligible, but will be included later on for completeness.
Throughout this work we shall 
neglect the contribution from the QCD penguin
operators $Q_3,\ldots, Q_6$, which enter at ${\cal O}(\alpha_s)$ and
are further suppressed by very small Wilson coefficients.
We note that to ${\cal O}(\alpha_s)$ the matrix element of $Q_2$
is zero because of its colour structure.
The amplitude for $\bar B\to K^*\gamma$ then reads
($\langle Q_i\rangle\equiv\langle K^*\gamma|Q_i|\bar B\rangle$)
\begin{equation}\label{abkgam718}
A(\bar B\to K^*\gamma)=\frac{G_F}{\sqrt{2}}\lambda_c^{(s)}
\left(C_7\langle Q_7\rangle +C_1\langle Q^c_1\rangle + 
C_8\langle Q_8\rangle \right)
\end{equation}
The leading-order matrix element $\langle Q_7\rangle$ is given in
(\ref{q7f1f2}) with $V\equiv K^*$. At subleading order in $\alpha_s$
the matrix elements $\langle Q_{1,8}\rangle$ need to be computed
from the diagrams in Figs. \ref{fig:qit1} -- \ref{fig:qit2}.
The result for the diagrams in Figs. \ref{fig:qit1} and \ref{fig:q8t1},
which enter the hard-scattering kernels $T^I_{1}$, $T^I_{8}$,
can be infered from \cite{GHW}. In these papers the diagrams were
computed to obtain the matrix elements for the inclusive mode
$b\to s\gamma$ at next-to-leading order. In this context
Figs. \ref{fig:qit1} and \ref{fig:q8t1} represented the virtual
corrections to the inclusive matrix elements of $Q_1$ and $Q_8$. 
In our case they determine the kernels $T^I_{1}$ and $T^I_{8}$.
As required for the consistency of the factorization formula these
corrections must be dominated by hard scales $\sim m_b$ and hence
must be infrared finite. This is indeed the case.
Re-interpreted as the perturbative hard-scattering kernels
for the exclusive process, the results from \cite{GHW} imply
\begin{equation}\label{q1me1}
\langle Q_{1,8}\rangle^I=\langle Q_7\rangle 
\frac{\alpha_s C_F}{4\pi} G_{1,8}
\end{equation}
where $C_F=(N^2-1)/(2N)$, with $N=3$ the number of colours, and
\begin{eqnarray}\label{G1}
G_1(s_c) &=& -\frac{104}{27}\ln\frac{\mu}{m_b}+ g_1(s_c) \\
\label{G8}
G_8 &=& \frac{8}{3}\ln\frac{\mu}{m_b} + g_8
\end{eqnarray}
\begin{eqnarray}\label{g1s}
g_1(s) &=& -\frac{833}{162}-\frac{20i\pi}{27} 
            +\frac{8\pi^2}{9} s^{3/2} \nonumber\\
        && {}+\frac{2}{9} \bigg[ 48+30i\pi-5\pi^2-2i\pi^3 -36\zeta (3) 
          +\left( 36+6i\pi-9\pi^2\right)\ln s \nonumber\\
        && \qquad\;\;\, +\left( 3+6i\pi\right) \ln^2\! s+\ln^3\! s \bigg] s 
           \nonumber\\
        && {}+\frac{2}{9} \bigg[ 18+2\pi^2 -2i\pi^3 
         +\left( 12-6\pi^2 \right)\ln s +6i\pi\ln^2\! s+\ln^3\! s\bigg] s^2 
           \nonumber\\
        && {}+\frac{1}{27} \bigg[ -9+112 i\pi-14\pi^2
          +\left(182-48i\pi\right)\ln s-126\ln^2\! s\bigg] s^3 \\
\label{g8}
g_8 &=& \frac{11}{3}-\frac{2\pi^2}{9}+\frac{2i\pi}{3}
\end{eqnarray}
Here we denote
\begin{equation}\label{sdef}
s_c=\frac{m^2_c}{m^2_b}
\end{equation}

We now turn to the mechanism where the spectator participates
in the hard scattering.

\begin{figure}[t]
   \epsfysize=10cm
   \epsfxsize=12cm
   \centerline{\epsffile{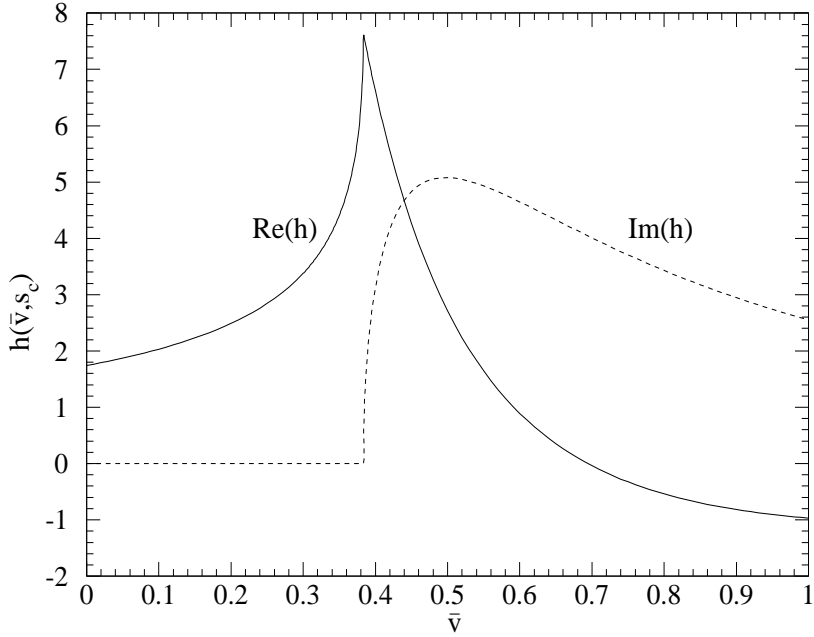}}
\caption{The hard-scattering kernel $h(\bar v,s_c)$ as a function
of $\bar v$. \label{fig:hcvbar}}
\end{figure}

To find the correction for $\langle Q_1\rangle$ we compute
the first diagram in Fig. \ref{fig:qit2}, using the
light-cone projectors in (\ref{phi12def}) and (\ref{phiperpdef}).
We obtain
\begin{equation}\label{q1me2}
\langle Q_1\rangle^{II}=\langle Q_7\rangle 
\frac{\alpha_s(\mu_h) C_F}{4\pi} H_1(s_c)
\end{equation}
with
\begin{equation}\label{h1s}
H_1(s)=-\frac{2\pi^2}{3 N}\frac{f_B f^\perp_V}{F_V m^2_B}
\int^1_0 d\xi\frac{\Phi_{B1}(\xi)}{\xi}\int^1_0 dv\, h(\bar v,s)
\Phi_\perp(v)
\end{equation}
The hard-scattering function $h(u,s)$ is given by
\begin{equation}\label{hus}
h(u,s)=\frac{4s}{u^2}\left[ 
L_2\!\left(\frac{2}{1-\sqrt{\frac{u-4s+i\varepsilon}{u}}}\right)+
L_2\!\left(\frac{2}{1+\sqrt{\frac{u-4s+i\varepsilon}{u}}}\right)\right]
-\frac{2}{u}
\end{equation}
$L_2$ is the dilogarithmic function
\begin{equation}\label{dilog}
L_2(x)=-\int^x_0 dt\frac{\ln(1-t)}{t}
\end{equation}
The function $h(u,s)$ is real for $u\leq 4s$ and develops an
imaginary part for $u > 4s$. At small values of $u$ it has the
expansion
\begin{equation}\label{husexp}
h(u,s)=\frac{1}{6s}+\frac{1}{45 s^2}u+
{\cal O}\left(\frac{u^2}{s^3}\right)
\end{equation}
It is also regular for $s\to 0$
\begin{equation}\label{hus0}
h(u,0)=-\frac{2}{u}
\end{equation}
The function $h(\bar v,s_c)$ is displayed in Fig. \ref{fig:hcvbar}.

The correction to $\langle Q_8\rangle$ from the hard spectator
interaction comes from the second diagram in Fig. \ref{fig:qit2}.
One finds 
\begin{equation}\label{q8me2}
\langle Q_8\rangle^{II}=\langle Q_7\rangle 
\frac{\alpha_s(\mu_h) C_F}{4\pi} H_8
\end{equation}
where
\begin{equation}\label{h8}
H_8=\frac{4\pi^2}{3 N}\frac{f_B f^\perp_V}{F_V m^2_B}
\int^1_0 d\xi\frac{\Phi_{B1}(\xi)}{\xi}
\int^1_0 dv\frac{\Phi_\perp(v)}{v}
\end{equation}

The gluons in Fig. \ref{fig:qit2} transfer a momentum of order 
$\mu_h \sim \sqrt{\Lambda_{QCD} m_b}$. 
Therefore we set $\alpha_s=\alpha_s(\mu_h)$ in (\ref{q1me2}) and
(\ref{q8me2}). For our numerical analysis we shall use 
$\mu_h=\sqrt{\Lambda_h \mu}$ with $\Lambda_h=0.5$ GeV and 
$\mu={\cal O}(m_b)$.

Finally, we can combine these results and write,
including also the up-quark contribution,
\begin{equation}\label{abkgam}
A(\bar B\to K^*\gamma)=\frac{G_F}{\sqrt{2}}
\left[\sum_p \lambda_p^{(s)}\, a^p_7(K^*\gamma)\right] 
\langle K^*\gamma|Q_7|\bar B\rangle
\end{equation}
where, at NLO
\begin{eqnarray}\label{a7vgam}
a^p_7(V\gamma) = C_7 &+& \frac{\alpha_s(\mu) C_F}{4\pi}
\left( C_1(\mu) G_1(s_p)+ C_8(\mu) G_8\right) \nonumber \\  
  &+& \frac{\alpha_s(\mu_h) C_F}{4\pi} 
  \left( C_1(\mu_h) H_1(s_p)+ C_8(\mu_h) H_8\right)
\end{eqnarray}
Here the NLO expression for $C_7$ has to be used, while the leading
order values are sufficient for $C_1$ and $C_8$. The explicit formulas for the 
Wilson coefficients can be found in \cite{CMM}.

The scale dependence of the matrix element of $Q_7$ is reflected in the 
running of the product of $b$-quark mass and form factor which is 
explicitely given as
\begin{equation}\label{runmbff}
\left( m_b \cdot F_V \right)[\mu] = \left( m_b \cdot F_V \right)[m_b]
\left(1-\frac{\alpha_s(\mu)}{4\pi} 8C_F \ln\frac{\mu}{m_b}\right)
\end{equation}
This dependence on the scale $\mu$ has to be taken into account
when the residual scale dependence of physical quantities
is investigated.

From the amplitude in (\ref{abkgam}) the branching ratio is obtained as
\begin{equation}\label{brbkgam}
B(\bar B\to K^*\gamma)=\tau_B\frac{G^2_F \alpha m^3_B m^2_b}{32\pi^4}
\left(1-\frac{m^2_{K^*}}{m^2_B}\right)^3 
|\sum_p \lambda_p^{(s)}\, a^p_7(K^*\gamma)|^2\, |F_{K^*}|^2
\end{equation}

\section{$B\to\rho\gamma$}\label{sec:Brhogam}

For the decay $\bar B\to\rho\gamma$ both sectors of the effective 
Hamiltonian have the same order of magnitude and have to be included. 
It is straightforward to translate the expressions from the previous 
section to this case. 
The amplitude can be written as
\begin{equation}\label{abrgam}
A(\bar B\to \rho\gamma)=\frac{G_F}{\sqrt{2}}
\left[\sum_p \lambda_p^{(d)}\, a^p_7(\rho\gamma) \right]
\langle \rho\gamma|Q_7|\bar B\rangle
\end{equation}
where $a^p_7(V\gamma)$ is given in (\ref{a7vgam}).
The branching fraction becomes
\begin{equation}\label{brbrgam}
B(\bar B\to \rho\gamma)=\tau_B\frac{G^2_F \alpha m^3_B m^2_b}{32\pi^4}
\left(1-\frac{m^2_\rho}{m^2_B}\right)^3 
|\sum_p \lambda_p^{(d)}\, a^p_7(\rho\gamma)|^2\, c^2_\rho |F_\rho|^2
\end{equation}
The rate for the CP-conjugated mode $B\to\rho\gamma$ 
is obtained by replacing $\lambda_p^{(d)} \to \lambda_p^{(d)*}$. 
We may then consider the CP asymmetry
\begin{equation}\label{acpbrgdef}
{\cal A}_{CP}(\rho\gamma)=
\frac{\Gamma(B\to\rho\gamma)-\Gamma(\bar B\to\rho\gamma)}{
      \Gamma(B\to\rho\gamma)+\Gamma(\bar B\to\rho\gamma)}
\end{equation}


A non-vanishing CP asymmetry appears at ${\cal O}(\alpha_s)$ only. 
Expanding ${\cal A}_{CP}$ in $\alpha_s$ and using the improved Wolfenstein 
parametrization for the CKM elements we obtain
\begin{eqnarray}\label{acpbrgamsimp}
{\cal A}_{CP}(\rho\gamma) &=&
\frac{2\,{\rm Im}\lambda^{(d)*}_u\lambda^{(d)}_c}{|\lambda^{(d)}_t|^2}
\, 
\frac{{\rm Im}a^{u*}_7 a^c_7}{|C_7|^2} \\
&=& -\frac{2\bar{\eta}}{(1-\bar\rho)^2+\bar{\eta}^2}
\frac{\alpha_s C_F}{4\pi}\, \frac{C_1}{C_7}\,
{\rm Im}\left(G_1(s_c)+H_1(s_c)-G_1(0)-H_1(0)\right)\nonumber
\end{eqnarray}

As we have discussed above, weak annihilation from the
leading operators $Q_{1,2}$ contributes to the $\bar B\to\rho\gamma$ 
amplitude only at ${\cal O}(\Lambda_{QCD}/m_b)$, but is enhanced
by large Wilson coefficients. These effects can be calculated in
QCD factorization. To lowest order in $\alpha_s$ we obtain
from the diagram in Fig. \ref{fig:ann}, using the projection
in (\ref{fvmveta})
\begin{eqnarray}\label{aannrho}
A_{ann}(B^-\to\rho^-\gamma) &=& 
\frac{G_F}{\sqrt{2}}\lambda^{(d)}_u\, a_1 b_u \,
\langle\rho^-\gamma|Q_7|B^-\rangle \\
A_{ann}(\bar B^0\to\rho^0\gamma) &=& 
\frac{G_F}{\sqrt{2}}\lambda^{(d)}_u\, a_2 b_d \,
\langle\rho^0\gamma|Q_7|\bar B^0\rangle 
\end{eqnarray}
Here
\begin{equation}\label{ac12}
a_{1,2}=C_{1,2}+\frac{1}{N} C_{2,1}
\end{equation}
(evaluated in leading logarithmic approximation) and
\begin{equation}\label{bubd}
b_u=\frac{2\pi^2 Q_u f_B f_\rho m_\rho}{F_\rho m^2_B \lambda_B}
\ \ \  b_d=\frac{1}{2}b_u
\end{equation}
Recalling that $F_\rho\sim m^{-3/2}_b$, $f_B\sim m^{-1/2}_b$
in the heavy-quark limit, we note that $b_u\sim \Lambda_{QCD}/m_b$.
This shows explicitly the power suppression of weak annihilation. 
The ratio of $b_d$ to $b_u$ is $-Q_d/Q_u=1/2$, where the minus sign
comes from the relative sign between the up-quark and down-quark 
components of the $\rho^0$ wave function (only the up-quarks
produce the $\rho^0$ in the annihilation process, while only the
down-quarks are relevant in $\langle Q_7\rangle$).

The annihilation components are included in the decay
amplitudes by substituting
\begin{eqnarray}\label{a7ba}
a^u_7 &\to& a^u_7 + b_u a_1 \ \ \ \ {\rm for}\ \ \ \ B^-\to\rho^-\gamma \\
a^u_7 &\to& a^u_7 + b_d a_2 \ \ \ \ {\rm for}\ \ \ \ \bar B^0\to\rho^0\gamma 
\end{eqnarray}

The annihilation contribution for $B^-\to K^{*-}\gamma$ is similar
to the one for $B^-\to\rho^-\gamma$, with obvious replacements
$\rho\to K^*$ in $b_u$. Its impact is very small for $B^-\to K^{*-}\gamma$ 
because of the strong CKM suppression of the up-quark sector.
It will be included in our numerical analysis for definiteness.
Within our approximations weak annihilation gives no contribution
to $\bar B^0\to \bar K^{*0}\gamma$.

We conclude this section with a few general remarks on the role
of power corrections in $B\to V\gamma$ decays.
Let us first repeat that the annihilation effect from operator 
$Q_{1}$ gives a numerically important power correction, because
it comes with a relative enhancement factor of 
$|C_1/C_7|\sim 3$. This leads to a $30\%$ correction in the
amplitude of the charged mode $B^-\to\rho^-\gamma$.
Note that the parameter $b_u$ in (\ref{bubd}), describing the
generic effect of the annihilation term, is only about $10\%$,
which is consistent with a $\Lambda_{QCD}/m_b$ correction of
canonical size\footnote{After the preprint version of the
present article had appeared, further isospin-breaking power corrections
were investigated in \cite{KN}. A sizeable effect of $11\%$
from penguin annihilation related to $Q_6$ was identified.
This is larger than naively expected, but still substantially
smaller than the effect of $b_u$ (for the $\rho$ channel),
which is presumably the largest power correction. The $Q_6$
annihilation contribution is still calculable, while other
terms of the same order are much smaller numerically.}.
In contrast to nonleptonic modes with pseudoscalar mesons
in the final state, no chirally enhanced power corrections
from the light-cone expansion of meson wave functions arise in our case. 
Finally, power corrections can also come from the loop
effects with up- and  charm quarks, whose leading-power
contributions were computed in (\ref{q1me2}).
These power corrections correspond to the region of
integration where the gluon becomes soft, that is 
$\bar v={\cal O}(\Lambda_{QCD}/m_b)$. Their contribution
is nonperturbative and cannot be calculated in the hard-scattering
formalism. Nevertheless, the expression in (\ref{q1me2}) can be used
to read off the scaling behaviour of these power corrections
in the heavy-quark limit. For the charm loop the kernel approaches
a constant $\sim m^2_b/m^2_c$ in the endpoint region
as shown in (\ref{husexp}). Taking into account the linear
endpoint suppression of the wave function $\Phi_\perp$, the integral in 
(\ref{q1me2}) over the region $\bar v\sim \Lambda_{QCD}/m_b$ 
therefore contributes a term of order 
$(\Lambda_{QCD}/m_b)^2\times (m_b/m_c)^2 = (\Lambda_{QCD}/m_c)^2$.
It is interesting that we thus recover the power behaviour of soft
contributions in the charm sector first pointed out in \cite{VOL}.
This was discussed for the inclusive decay $b\to s\gamma$ in
\cite{VOL,BIR} and for the exclusive mode $B\to K^*\gamma$ in
\cite{KRSW}. Numerically this correction is very small ($\sim 3\%$
in the decay rate).
A similar consideration applies to the up-quark sector. In this case
the endpoint behaviour of the kernel is singular (\ref{hus0}),
which now leads to a linear power suppression of the form
$\Lambda_{QCD}/m_b$. This coincides with the scaling
behaviour derived in \cite{BIR} in the context of the
inclusive process.

In view of these remarks, we see no indication for large
uncalculable power corrections. Further studies on this
issue are desireable, but would go beyond the scope of the
present analysis.

\section{Phenomenology}\label{sec:phen}

In this section we present numerical results for various
observables of interest for the phenomenology of 
$B\to V\gamma$ decays, based on the NLO QCD expressions derived
in this paper.
Our choice of input parameters is summarized in Table \ref{tab:input}.
Our default choice for the CKM angle $\gamma$ is $58$ deg.
\begin{table}[p]
\caption[]{Summary of input parameters.}
\label{tab:input}
\renewcommand{\arraystretch}{1.2}
\begin{center}
\begin{tabular*}{140mm}{@{\extracolsep\fill}|c|c|c|c|c|c|}
\hline\hline
\multicolumn{6}{|c|}{CKM parameters and coupling constants}\\
\hline
$V_{us}$ & $V_{cb}$ & $\left|V_{ub}/V_{cb}\right|$ & 
$\Lambda_{\overline{MS}}^{(5)}$ & $\alpha$ & $G_F$\\
\hline
0.22 & 0.041 & $0.085 \pm 0.025$ & 225 MeV & 1/137 & $1.166 \times 10^{-5} 
\mbox{GeV}^{-2}$\\
\hline
\end{tabular*}

\begin{tabular*}{140mm}{@{\extracolsep\fill}|c|c|c|c|c|}
\hline
\multicolumn{5}{|c|}{Parameters related to the $B$ mesons}\\
\hline
$m_B$ & $f_B$ & $\lambda_B$ & $\tau_{B^+}$ & $\tau_{B^0}$\\
\hline
5.28 GeV & 180 MeV & $(350 \pm 150)$ MeV & 1.65 ps & 1.56 ps\\
\hline
\end{tabular*}

\begin{tabular*}{140mm}{@{\extracolsep\fill}|c|c|c|c|c|c|}
\hline
\multicolumn{6}{|c|}{Parameters related to the $K^*$ meson \cite{BB98}}\\
\hline
$F_{K^*}$ & $f_{K^*}^\perp$ & $m_{K^*}$ & $\alpha_1^{K^*}$ & 
$\alpha_2^{K^*}$ & $f_{K^*}$ \\
\hline
$0.38 \pm 0.06$ & 185 MeV & 894 MeV & 0.2 & 0.04 & 230 MeV\\
\hline\hline
\multicolumn{6}{|c|}{Parameters related to the $\rho$ meson \cite{BB98}}\\
\hline
$F_\rho$ & $f_\rho^\perp$ & $m_\rho$ & $\alpha_1^\rho$  & $\alpha_2^\rho$ & 
$f_\rho$\\
\hline
$0.29 \pm 0.04$ & 160 MeV & 770 MeV & 0 & 0.2 & 200 MeV\\
\hline
\end{tabular*}

\begin{tabular*}{140mm}{@{\extracolsep\fill}|c|c|c|c|}
\hline
\multicolumn{4}{|c|}{Quark and W-boson masses}\\
\hline
$m_b(m_b)$ & $m_c(m_b)$ & $m_{t,\mbox{pole}}$ & $M_W$\\
\hline
4.2 GeV & $(1.3 \pm 0.2)$ GeV & 174 GeV & 80.4 GeV\\
\hline\hline
\end{tabular*}
\end{center}
\end{table}

We begin with the numerical result for the NLO QCD coefficient 
$a_7^c(K^*\gamma)$ as a typical example. For central values of all 
parameters, at $\mu=m_b$, and displaying separately the size of the various 
correction terms, we find
\begin{eqnarray}\label{numa7cK}
a_7^c(K^* \gamma) &=& \begin{array}[t]{cccc}
-0.3221 & +0.0113 & -0.0820-0.0147i & -0.0144-0.0109i\\
C_7^{LO} & \Delta C_7^{NLO} & T^I_{1,8}\mbox{-contribution} & 
  T^{II}_{1,8}\mbox{-contribution}\\
\end{array} \nonumber\\
&=& -0.4072 -0.0256i.
\end{eqnarray}
\begin{table}[p]
\label{tab:output}
\caption[]{Predictions for branching ratios and CP asymmetries with the 
errors from the individual input uncertainties.}
\renewcommand{\arraystretch}{1.1}
\begin{center}
\begin{tabular}{|l|c|c|c|}
\hline\hline
   & $B(\bar{B}^0 \to \bar{K}^{*0} \gamma)[10^{-5}]$    
   & 
     $B(B^- \to \rho^- \gamma)[10^{-6}]$        & 
${\cal A}_{CP}(\rho^\pm\gamma)[\%]$\\
\hline\hline
central                         & 7.09                                  
  & 1.58                                        & 9.89\\
\hline\hline
$F_{K^*}$                       & +2.32/$-$1.99                         
  &   --    &  --  \\
\hline
$F_{\rho}$                      &   --                                     
  & +0.42/$-$0.37                               & +0.12/$-$0.13\\
\hline
$\mu$                           & +0.67/$-$1.00                         
  & +0.07/$-$0.19                               & +5.25/$-$2.40\\
\hline
$\lambda_B$                     & +0.38/$-$0.15                         
  & +0.26/$-$0.09                               & +0.35/$-$0.28\\
\hline
$m_c$                           & +0.38/$-$0.43                         
  & +0.12/$-$0.12                               & +1.40/$-$1.52\\
\hline
$\left|V_{ub}/V_{cb}\right|$    & +0.05/$-$0.05                         
  & +0.15/$-$0.13                               & +3.48/$-$3.30\\
\hline\hline
\end{tabular}
\end{center}
\end{table}
We note a sizable enhancement of the leading order value, dominated by the 
$T^I$-type correction. This feature was already observed in the context of 
the inclusive case in \cite{GHW}. A complex phase is generated at NLO, 
where the $T^I$-corrections and the hard-spectator interactions ($T^{II}$) 
yield comparable effects.

\begin{figure}[t]
\epsfig{figure=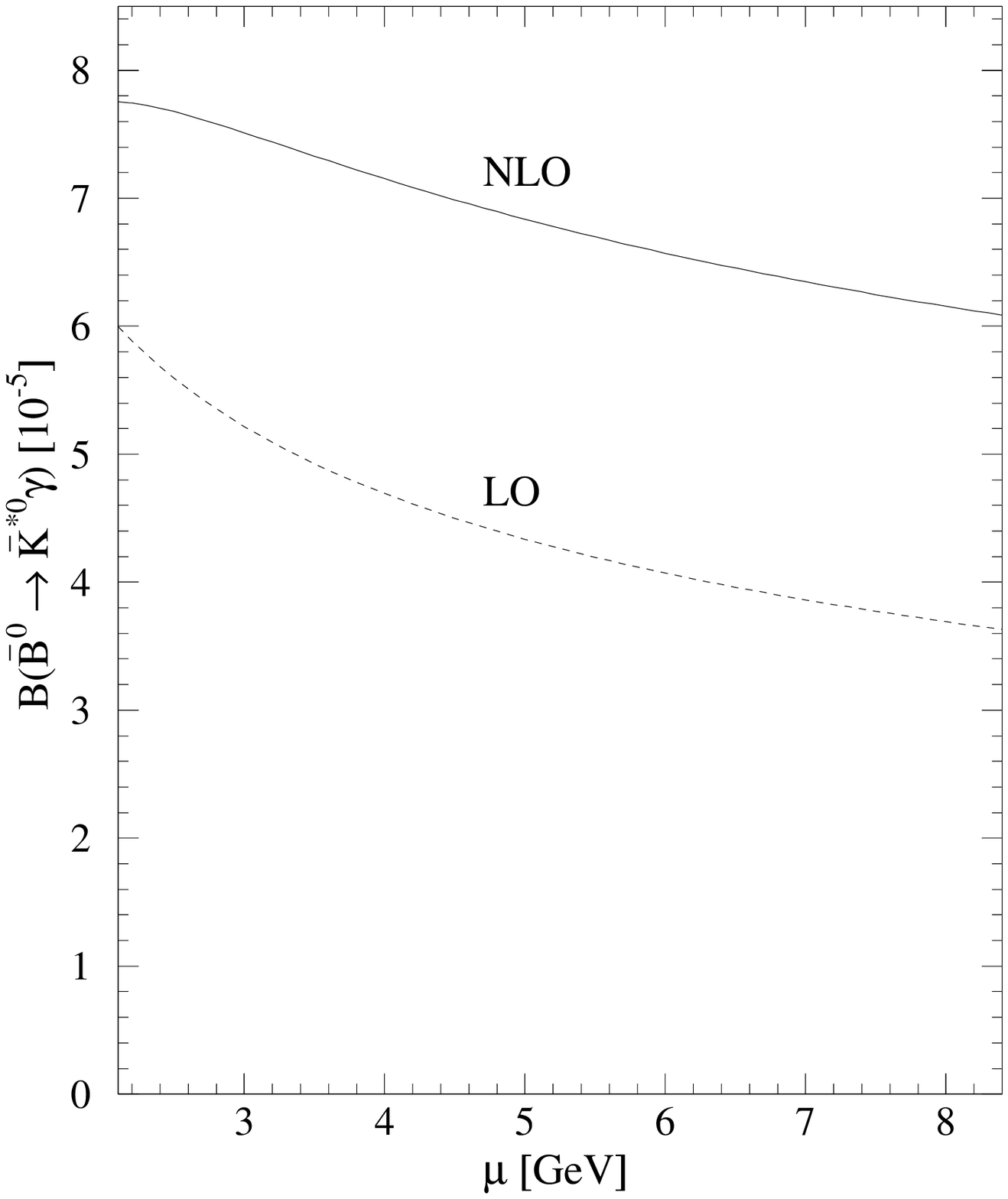,width=7.2cm,height=8.0cm}
\epsfig{figure=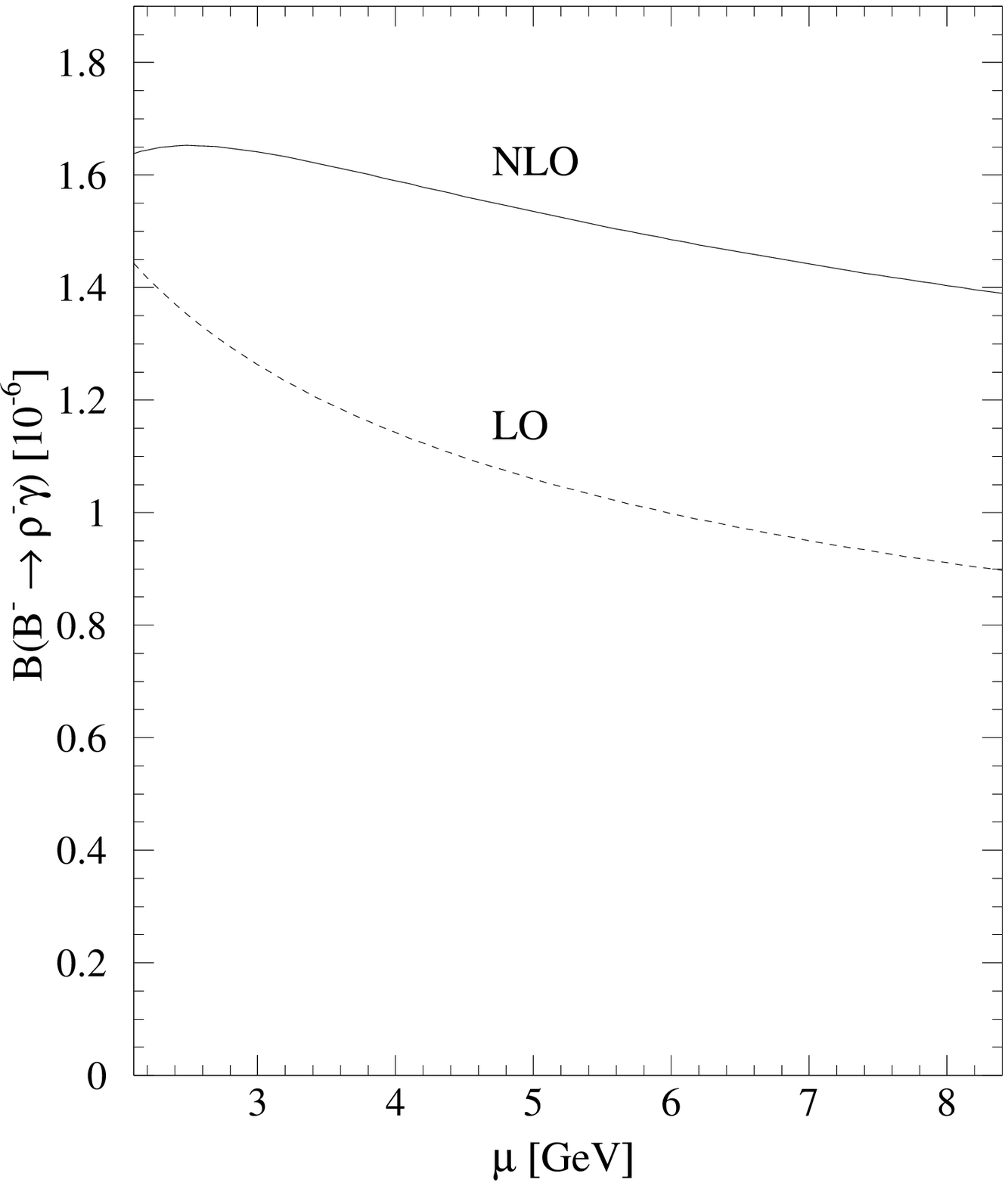,width=7.2cm,height=8.0cm}
\caption{Dependence of the branching fractions 
$B(\bar{B}^0 \to \bar{K}^{*0} \gamma)$ and $B(B^- \to \rho^- \gamma)$ 
on the renormalization scale $\mu$ at leading and
next-to-leading order.}
\label{fig:bkrhomu}
\end{figure}

The net enhancement of $a_7$ at NLO leads to a corresponding enhancement 
of the branching ratios, for fixed value of the form factor. This is 
illustrated in Fig. \ref{fig:bkrhomu}, where we show the residual scale 
dependence for $B(\bar{B}\to \bar{K}^{*0}\gamma)$ and $B(B^-\to\rho^-\gamma)$ 
at leading and next-to-leading order.

The sensitivity of the $\bar{B}\to \bar{K}^{*0}\gamma$ and 
$B^-\to\rho^-\gamma$ branching ratios, and of the CP asymmetry 
${\cal A}_{CP}(\rho\gamma)$ to variations in the relevant input parameters 
are summarized in Table 2. The uncertainty of the branching fractions is 
currently dominated by the form factors $F_{K^*}$, $F_\rho$. The values for 
$B(B^-\to K^{*-}\gamma)$ are very close to those for 
$B(\bar{B}\to \bar{K}^{*0}\gamma)$ with the main shift coming
from the different lifetimes. For central input parameters we find
$B(B^-\to K^{*-}\gamma)=7.45\cdot 10^{-5}$ compared to
$B(\bar{B}\to \bar{K}^{*0}\gamma)=7.09\cdot 10^{-5}$.
The CP asymmetry ${\cal A}_{CP}(K^*\gamma)$ is typically
$-0.5\%$.

Taking the sizable uncertainties into 
account, the results for $B\to K^*\gamma$ in Table 2 are compatible with the 
experimental measurements in (\ref{b0kgamex}) and (\ref{bpkgamex}), 
even though the central theoretical values appear to be somewhat high.

\begin{figure}[t]
   \epsfysize=10cm
   \epsfxsize=10cm
   \centerline{\epsffile{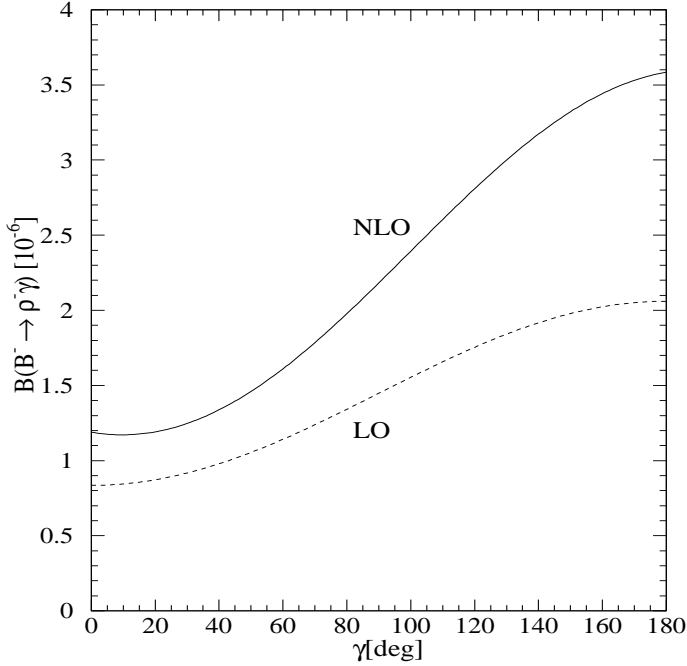}}
\caption{The branching fraction $B(B^- \to \rho^- \gamma)$ as a function
of the CKM angle $\gamma$ at leading and next-to-leading order. 
\label{fig:brhogamma}}
\end{figure}

\begin{figure}[t]
   \epsfysize=10cm
   \epsfxsize=12cm
   \centerline{\epsffile{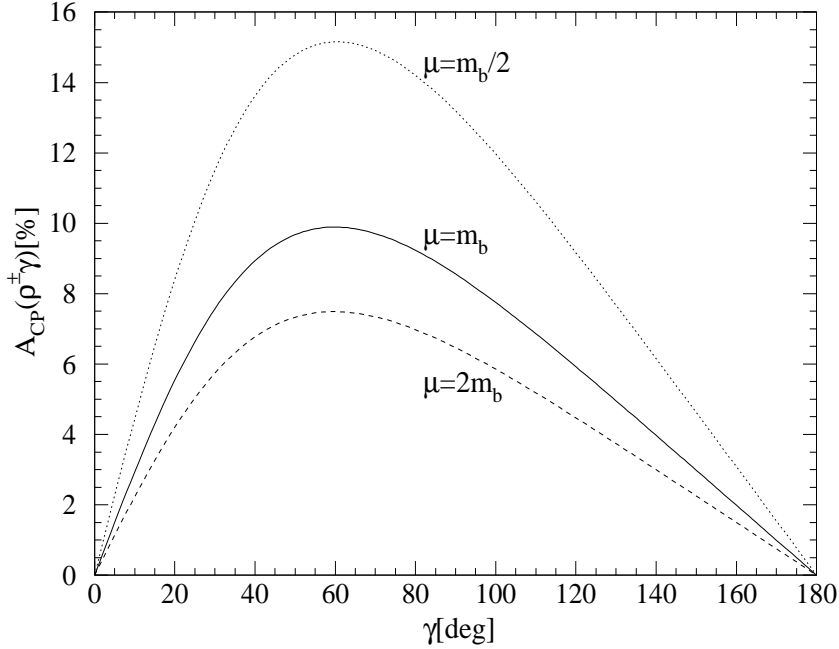}}
\caption{The CP asymmetry ${\cal A}_{CP}(\rho\gamma)$ 
as a function of the CKM angle $\gamma$ for three
values of the renormalization scale $\mu=m_b/2$, $m_b$
and $2 m_b$. \label{fig:cpasymrho}}
\end{figure}

The direct CP asymmetry, which is substantial for the $\rho\gamma$ modes, 
is much less dependent on the form factors. Here the largest theoretical 
uncertainty comes from the scale dependence. This is to be expected because 
the direct CP asymmetry is proportional to the perturbative strong phase 
difference, which arises at $\ord(\as)$. Unknown power corrections 
could have some impact on the prediction. 
$B\to\rho\gamma$ also depends sensitively on fundamental CKM 
parameters, such as $\left|V_{ub}/V_{cb}\right|$ and $\gamma$,
and can thus in principle serve to constrain the latter quantities 
once measurements become available.
This is further illustrated in Figs. \ref{fig:brhogamma} and 
\ref{fig:cpasymrho}, where the dependence on $\gamma$ is shown for 
$B(B^-\to\rho^-\gamma)$ and ${\cal A}_{CP}(\rho\gamma)$, respectively.

\begin{figure}[t]
   \epsfysize=10cm
   \epsfxsize=12cm
   \centerline{\epsffile{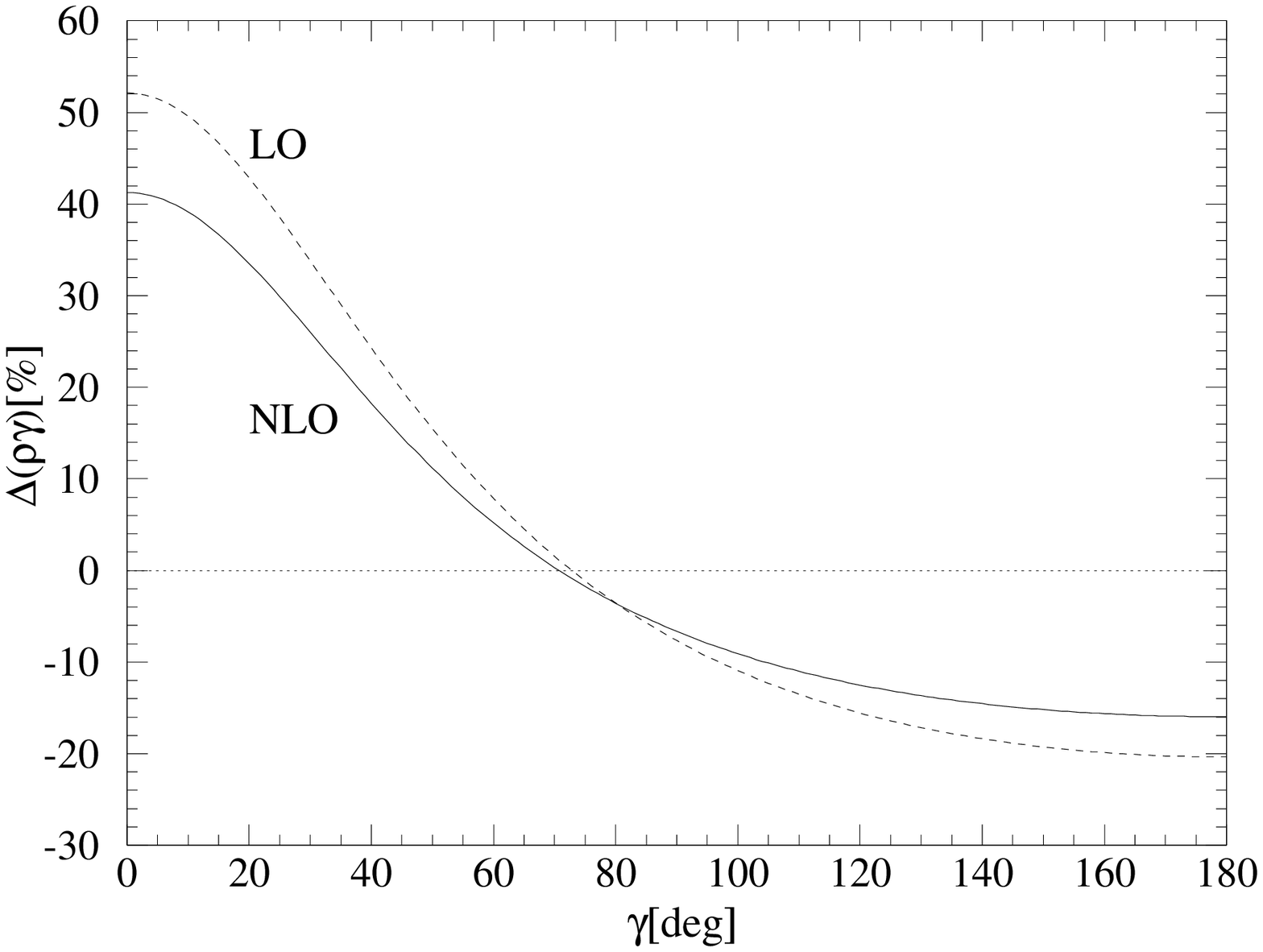}}
\caption{The isospin-breaking asymmetry $\Delta(\rho\gamma)$ 
as a function of the CKM angle $\gamma$ at leading and
next-to-leading order. \label{fig:isodeltarho}}
\end{figure}

Further interesting observables are the isospin breaking quantities
\begin{eqnarray}
\label{isobreak}
\Delta_{+0} &=& 
\frac{\Gamma(B^+\to\rho^+\gamma)}{2\Gamma(B^0\to\rho^0\gamma)}-1\\
\Delta_{-0} &=& 
\frac{\Gamma(B^-\to\rho^-\gamma)}{2\Gamma(\bar{B}^0\to\rho^0\gamma)}-1\\
\Delta(\rho\gamma) &=& \frac{\Delta_{+0}+\Delta_{-0}}{2}
\end{eqnarray}
Within our approximations, isospin breaking is generated by weak 
annihilation. Isospin breaking was already discussed in \cite{AHL}, partially 
including NLO corrections. Our results, including the power-suppressed 
annihilation effects at leading order in QCD, together with the complete 
NLO expressions at leading power, are displayed in Fig. \ref{fig:isodeltarho}.
We remark that our sign of $\Delta(\rho\gamma)$ differs
from the one found in \cite{AHL}.

Another application of our results concerns an estimate of
U-spin breaking effects in $B\to V\gamma$ decays.
U-spin symmetry, the symmetry of strong interactions under exchange
of $d$ and $s$ quarks, has been advocated as a tool to control
hadronic uncertainties in tests of the Standard Model \cite{GRG,HM,RF}.
Defining
\begin{eqnarray}\label{dbbkr}
\Delta B(B\to K^*\gamma) &=& 
     B(B^+\to K^{*+}\gamma) - B(B^-\to K^{*-}\gamma) \\
\Delta B(B\to \rho\gamma) &=& 
     B(B^+\to \rho^{+}\gamma) - B(B^-\to \rho^{-}\gamma) 
\end{eqnarray}
the quantity
\begin{equation}\label{dbb}
\Delta B(B\to K^*\gamma)+\Delta B(B\to \rho\gamma) \equiv 0
\end{equation}
in the limit of U-spin symmetry and within the Standard Model.
This has been discussed in \cite{GRG} and was 
considered in more detail in \cite{HM}.
Using our expressions and central values for all
parameters we find
\begin{eqnarray}\label{dbbkrnum}
\Delta B(B\to K^*\gamma) &=& -7\cdot 10^{-7} \\
\Delta B(B\to \rho\gamma) &=& 4\cdot 10^{-7} 
\end{eqnarray}
where we have chosen the CKM angle $\gamma=\pi/2$, which
maximises the effects. The two quantities indeed have opposite
signs, but their sum only partly cancels, leaving a U-spin
breaking remainder of $-3\cdot 10^{-7}$. This effect is
almost entirely due to the difference $(F_{K^*}-F_\rho)$.
For form-factor values different from those in Table 1,
the U-spin breaking effect would approximately be rescaled 
proportional to $(F_{K^*}-F_\rho)$.
For our choice the sum of the two asymmetries in (\ref{dbb})
is of the same order of magnitude as the individual asymmetries.
This example quantifies the limitations of the relation
(\ref{dbb}) as a Standard Model test.

\section{Conclusions}\label{sec:concl}

In this paper we have discussed a systematic and
model-independent framework for the exclusive radiative
decays $B\to V\gamma$ based on the heavy-quark limit.
This allowed us to compute the decay amplitudes for these modes
consistently at next-to-leading order in QCD.

An important conceptual aspect of this analysis is the interpretation
of loop contributions with charm and up quarks, which come from
leading operators in the effective weak Hamiltonian.
We have argued that these effects are calculable in terms of
perturbative hard-scattering functions and universal meson
light-cone distribution amplitudes. They are ${\cal O}(\alpha_s)$
corrections, but are leading power contributions in the
framework of QCD factorization. This picture is in contrast to the
common notion that considers charm and up-quark loop effects as
generic, uncalculable long-distance contributions.
Non-factorizable long-distance corrections may still exist, but
they are power-suppressed.

Another important feature of the NLO calculation are the strong 
interaction phases, which are calculable at leading power. They
play a crucial role for CP violating observables.

We have seen that weak-annihilation amplitudes are power-suppressed,
but can be numerically important for $B\to\rho\gamma$ because
they enter with large coefficients. These effects also turn out to
be calculable and were included in our phenomenological
discussion at leading order in QCD.

Finally, we have presented a numerical analysis of important
observables, including the branching ratios of
$B\to K^*\gamma$, $B\to\rho\gamma$, and CP asymmetries and 
isospin-breaking in $B\to\rho\gamma$ decays (see sec. \ref{sec:phen}).
Currently, the large uncertainties in the $B\to V$ form factors
are still an important limitation, but the situation can be
systematically improved. In particular, our approach allows for
a consistent perturbative matching of the nonperturbative form factor 
to the short-distance part of the amplitude.
Our formalism can also be applied to other radiative rare $B$ decays,
such as $B_s\to V\gamma$, $V=\phi$, $K^*$, or $B_d\to\omega\gamma$
(see e.g. \cite{GP} for a discussion of these modes).

The improved theoretical understanding of $B\to V\gamma$ decay
observables streng\-thens the motivation for still more detailed
experimental investigations, which will contribute
significantly to our knowledge of the flavour sector.

\section*{Acknowledgements}
While this paper was being written, we became aware of the work
in ref. \cite{BFS} where a similar subject is treated. 
We thank Martin Beneke for useful discussions and communications
on the results of \cite{BFS} prior to publication.  
Thanks are also due to Josip Trampeti\'c for discussions.
S.W.B. gratefully acknowledges financial support from the
Studienstiftung des deutschen Volkes and thanks the CERN Theory
Division for the kind hospitality.

\section*{Note Added}
After completion of this work we received the manuscript
\cite{AP}\footnote{In the meantime a substantially revised
version of this paper has appeared, where the points mentioned
here have been addressed.} 
discussing $B\to\rho\gamma$ decays at next-to-leading
order in QCD. The calculation uses the framework of \cite{BBNS1,BBNS2}
and of \cite{BF}, and is similar in spirit to the analysis presented
here. We disagree, however, with several of the main results given
in \cite{AP}. First, the expressions corresponding to
our functions $H_1(0)$ and $H_8$ are missing a factor of
$-4$ and $-1$, respectively, in \cite{AP}. Furthermore,
the charm contribution $H_1(s_c)$ is approximated by $H_1(0)$.
As can be seen from our results, this approximation is not
suitable. It gives a real part with the wrong sign and misses
a complex phase, which is important for ${\cal A}_{CP}(\rho\gamma)$
(changing the asymmetry in Table 2 from $9.9\%$ to $13.2\%$).
Finally, we find $H_1(s_c)$ to be free of infrared divergences,
also for non-vanishing charm-quark mass, contrary to \cite{AP}.
This is important conceptually, because such a divergence could not
be reabsorbed into the form factor and would signify a breakdown
of the factorization approach.

\vfill\eject
 

\begin{thebibliography}{99}

\bibitem{BLA}
F. Blanc [CLEO Collaboration], talk presented at the
XXXVIth Rencontres de Moriond, Electroweak Interactions and
Unified Theories, Les Arcs, France, March 2001.
\bibitem{TAY}
G. Taylor, [BELLE Collaboration], talk presented at the
XXXVIth Rencontres de Moriond, Electroweak Interactions and
Unified Theories, Les Arcs, France, March 2001.
\bibitem{BAR}
R.~Barate {\it et al.}  [ALEPH Collaboration],
Phys.\ Lett.\ B {\bf 429} (1998) 169.


\bibitem{COA}
T.~E.~Coan {\it et al.}  [CLEO Collaboration],
Phys.\ Rev.\ Lett.\ {\bf 84} (2000) 5283.
\bibitem{PUL}
T. Pulliam [BABAR Collaboration], talk presented at PHENO 2001,
Madison, Wisconsin, May 2001, BaBar-Talk-01/53.

 





\bibitem{BBM}
S.~Bertolini, F.~Borzumati and A.~Masiero,
Phys.\ Rev.\ Lett.\ {\bf 59} (1987) 180.

\bibitem{AGM}
A.~Ali, C.~Greub and T.~Mannel,
DESY-93-016, {\it To be publ. in Proc. of ECFA Workshop on the 
Physics of a B Meson Factory, Eds. R. Aleksan, A. Ali, 1993}.

\bibitem{BMMP}
A.~J.~Buras, M.~Misiak, M.~M\"u{}nz and S.~Pokorski,
Nucl.\ Phys.\ B {\bf 424} (1994) 374.

\bibitem{CFMRS}
M.~Ciuchini {\it et al.},
Phys.\ Lett.\ B {\bf 316} (1993) 127;
M.~Ciuchini, E.~Franco, L.~Reina and L.~Silvestrini,
Nucl.\ Phys.\ B {\bf 421} (1994) 41.




\bibitem{AY}
K.~Adel and Y.~Yao,
Phys.\ Rev.\ D {\bf 49} (1994) 4945;
C.~Greub and T.~Hurth,
Phys.\ Rev.\ D {\bf 56} (1997) 2934;
A.~J.~Buras, A.~Kwiatkowski and N.~Pott,
Nucl.\ Phys.\ B {\bf 517} (1998) 353.

\bibitem{GHW}
C.~Greub, T.~Hurth and D.~Wyler,
Phys.\ Rev.\ D {\bf 54} (1996) 3350;
A.~J.~Buras, A.~Czarnecki, M.~Misiak and J.~Urban,
hep-ph/0105160.



\bibitem{CMM}
K.~Chetyrkin, M.~Misiak and M.~M\"unz,
Phys.\ Lett.\ B {\bf 400} (1997) 206;
Erratum-ibid. B {\bf 425} (1998) 414.

\bibitem{MG}
M.~Misiak,
hep-ph/0002007;
C.~Greub,
hep-ph/9911348.


\bibitem{AAWGSW}
H.~H.~Asatrian, H.~M.~Asatrian and D.~Wyler,
Phys.\ Lett.\ B {\bf 470} (1999) 223; 
C.~Greub, H.~Simma and D.~Wyler,
Nucl.\ Phys.\ B {\bf 434} (1995) 39
[Erratum-ibid.\ B {\bf 444} (1995) 447]


\bibitem{BBNS1}
M.~Beneke, G.~Buchalla, M.~Neubert and C.~T.~Sachrajda,
Phys.\ Rev.\ Lett.\ {\bf 83} (1999) 1914.

\bibitem{BBNS2}
M.~Beneke, G.~Buchalla, M.~Neubert and C.~T.~Sachrajda,
Nucl.\ Phys.\ B {\bf 591} (2000) 313.


\bibitem{BB98}
P.~Ball and V.~M.~Braun,
Phys.\ Rev.\ D {\bf 58} (1998) 094016.

\bibitem{FS}
J.~M.~Flynn and C.~T.~Sachrajda,
hep-lat/9710057.


\bibitem{CLOPR}
J.~Charles {\it et al.},
Phys.\ Rev.\ D {\bf 60} (1999) 014001.
\bibitem{BF}
M.~Beneke and T.~Feldmann,
Nucl.\ Phys.\ B {\bf 592} (2001) 3.
\bibitem{BH}
G.~Burdman and G.~Hiller,
hep-ph/0011266.


\bibitem{BBL}
G.~Buchalla, A.~J.~Buras and M.~E.~Lautenbacher,
Rev.\ Mod.\ Phys.\ {\bf 68} (1996) 1125.

\bibitem{ABS}
A.~Ali, V.~M.~Braun and H.~Simma,
Z.\ Phys.\ C {\bf 63} (1994) 437.

\bibitem{BB96}
P.~Ball and V.~M.~Braun,
Phys.\ Rev.\ D {\bf 54} (1996) 2182.


\bibitem{GP}
B.~Grinstein and D.~Pirjol,
Phys.\ Rev.\ D {\bf 62} (2000) 093002.

\bibitem{KN}
A.~L.~Kagan and M.~Neubert,
hep-ph/0110078.

\bibitem{VOL}
M.~B.~Voloshin,
Phys.\ Lett.\ B {\bf 397} (1997) 275;
Z.~Ligeti, L.~Randall and M.~B.~Wise,
Phys.\ Lett.\ B {\bf 402} (1997) 178;
A.~K.~Grant et al.,
Phys.\ Rev.\ D {\bf 56} (1997) 3151.

\bibitem{BIR}
G.~Buchalla, G.~Isidori and S.~J.~Rey,
Nucl.\ Phys.\ B {\bf 511} (1998) 594.

\bibitem{KRSW}
A.~Khodjamirian et al.,
Phys.\ Lett.\ B {\bf 402} (1997) 167;




\bibitem{AHL}
A.~Ali, L.~T.~Handoko and D.~London,
hep-ph/0006175.


\bibitem{GRG}
M.~Gronau and J.~L.~Rosner,
Phys.\ Lett.\ B {\bf 500} (2001) 247;
M.~Gronau,
Phys.\ Lett.\ B {\bf 492} (2000) 297.
\bibitem{HM}
T.~Hurth and T.~Mannel,
hep-ph/0103331.
\bibitem{RF}
R.~Fleischer,
Phys.\ Lett.\ B {\bf 459} (1999) 306.




\bibitem{BFS}
M. Beneke, T. Feldmann and D. Seidel, 
RWTH Aachen preprint PITHA 01/05, hep-ph/0106067.

\bibitem{AP}
A.~Ali and A.~Y.~Parkhomenko,
hep-ph/0105302v1.



\end{thebibliography}
\end{document}